\documentclass[aps,preprint,nofootinbib]{revtex4}
\usepackage{graphicx}
\usepackage{epsf}
\usepackage{wrapfig}
\usepackage{epsfig}
\usepackage{sublabel}
\usepackage{subfigure}
\usepackage{epsfig}
\newcommand{\be}{\begin{eqnarray}}
\newcommand{\ee}{\end{eqnarray}}
\newcommand{\beq}{\begin{eqnarray}}
\newcommand{\eeq}{\end{eqnarray}}
\newcommand{\bey}{\begin{eqnarray}}
\newcommand{\eey}{\end{eqnarray}}

\newcommand{\sez}{\frac{d^{2}\sigma(q,\nu)}{d{\Omega}_2\,d{\nu}}}

\begin{document}
\title{A novel analysis of the effects of short range correlations in
inclusive lepton scattering off nuclei\footnote{ Extended version of
  a presentation given by CBM at the "Mini-symposium on Nuclear Structure at
  Short Distances" held within the "American Physical Society  April Meeting", Denver,
  Colorado (USA), $2-5$ May $2009$. \emph{Bulletin of the American Physical Society}, vol. $\textbf{54}$ n. $4$}}
\author{Chiara Benedetta Mezzetti and Claudio Ciofi degli Atti,}
\address{
Department of Physics, University of  Perugia and  Istituto
Nazionale di Fisica Nucleare,\\ Sezione di Perugia, Via A.Pascoli,
I-06100 Perugia, Italy }
\begin{abstract}
It is shown that, if inclusive lepton scattering off nuclei at high
 momentum transfer ($Q^2\gtrsim \, 1\,GeV^2$) is analyzed in terms
 of proper scaling variables,  useful information on
 Nucleon-Nucleon short range correlations in nuclei can be obtained.
The traditional approach to $Y$-scaling is critically
analyzed and a novel approach to $Y$-scaling, which incorporates
the effects from two- and three-nucleon correlations in nuclei, is illustrated.
\end{abstract}
 \maketitle
%
%
\section{Introduction}
New data on inclusive quasi elastic (q.e.) electron scattering off
nuclei, $A(e,e')X$, at high momentum transfer ($2.5 \lesssim Q^2
\lesssim 7.4\,GeV^2$) are under analysis at the Thomas Jefferson
National Accelerator Facility (JLab) \cite{Fomin}. Nowadays one of the aims
of the investigation of q.e. scattering off nuclei is to obtain
information on Nucleon-Nucleon (NN) short range correlations
(SRC); to this end various approaches are being pursued, such as
the investigation of the scaling behavior of the ratio of the inclusive
cross section $\sigma_2^A$ of heavy nuclei to that of $^2H$ and $^3He$ plotted
versus the Bjorken scaling variable $x_{Bj}$ \cite{ratioAD,Egyian}, or the
analysis of cross sections in terms of $Y$-scaling
\cite{CW}. The aim of this talk is to critically review these
analyses and propose a novel approach to $A(e,e')X$ processes
particularly suited to treat the effects of SRC.
%
%
\section{Cross section ratios: experimental results}
\begin{figure}[!hbp]
\centerline{\centerline{\epsfig{file=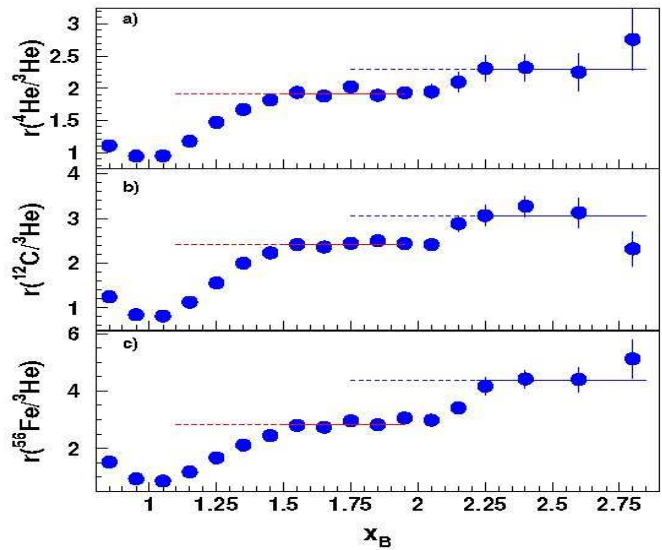,width=9cm,height=7.5cm}}}
\vskip -0.3cm \caption{The inclusive cross section ratio of $^{56}Fe$, $^{12}C$  and $^{4}He$ to $^{3}He$ \emph{vs.} the Bjorken scaling variable $x_B=Q^2/(2m_N\nu)\equiv x_{Bj}$. After Ref. \cite{Egyian}.}\label{Fig1}
\end{figure}
In Fig. \ref{Fig1}, the experimental cross section ratio
$\sigma_2^A/\sigma_2^{A'}\equiv r(A/A')$ plotted versus the Bjorken scaling variable $x_{Bj}$ is shown \cite{Egyian};
three distinct kinematical regions can be observed: i) the first
one, at $x_{Bj}\lesssim 1.5$, is due to the contribution of mean
field nucleons, and its shape is governed by the different
behaviour of the magnitude of the q.e. peak for different nuclei
(higher peaks for light nuclei and lower peaks for heavy nuclei);
ii) the second region, at $1.5 \lesssim x_{Bj}\lesssim 2$,
exhibits a plateaux, which is interpreted as due to two-nucleon
correlations (2NC); iii) the third region, at $2 \lesssim
x_{Bj}\lesssim 3$, seems to show a second plateaux, which is ascribed to three-nucleon correlations (3NC). Following the
original suggestion of Ref. \cite{FraStri}, the presence of these
plateaux can be viewed as evidence that two- and three-nucleon correlations in complex nuclei and in
$^3He$ differ only by a scale factor. It should however be pointed
out that no direct calculations of the cross section ratio shown
in Fig. \ref{Fig1} have been performed so far. These calculations would represent a relevant contribution towards the solution of the longstanding problem concerning the role played by SRC in nuclei. Here,
preliminary results of the calculation of the ratio $r(A/A')$ will be given. In order to
illustrate
the
basic ideas  of our approach \cite{ciocbm}, some general concepts of Y-scaling have to be recalled.
%
%
\section{Inclusive Lepton Scattering and $Y$-scaling}
In PWIA, the inclusive q.e. cross section can be written as follows
\cite{ciofi}
\be
 \sigma_2^A (q,\nu)\equiv \sez
 = F^A(q,\nu)\: K(q,\nu) \: \left[ Z\sigma_{ep} + N\sigma_{en} \right]
 \label{X-section}
\ee
where
\beq \label{Funzscala}
    F^A(q,\nu)=2\pi \int_{E_{min}}^{E_{max}(q,\nu)} dE \int_{k_{min}(q,\nu,E)}^{k_{max}(q,\nu,E)} k\:dk\: P^A(k,E)
\eeq is the nuclear structure function,
$\textbf{q}=\textbf{k}_1-\textbf{k}_2$ ($|\textbf{q}|=q$) and $\nu=\epsilon_1-\epsilon_2$ are the
three-momentum and energy transfers, $\sigma_{eN}$ is the elastic
electron cross section off a moving off-shell nucleon with
momentum $k\equiv |\textbf{k}|$ and removal energy $E$, $K(q,\nu)$
is a kinematical factor, and, eventually, $P^A(k,E)$ is the spectral
function. For ease of presentation, we will consider high values
of the momentum transfer, such that $E_{max}(q,\nu)$ and
$k_{max}(q,\nu,E)$ become very large, in which case, owing to the
rapid falloff of $P^A(k,E)$ with $k$ and $E$,  the replacement
$E_{max}=k_{max}=+\infty$ is justified. Without any loss of generality, we can
substitute the energy transfer $\nu$ with a generic "scaling variable" $Y=Y(q,\nu)$;
in this case, the "scaling function" (\ref{Funzscala}) can be
written as follows \cite{ciofi} \be
    F^A(q,Y)=f^A(Y)-B^A(q,Y)
\ee where the first term
\beq \label{longitudinal}
f^A(Y)=2\pi \int_{|Y|}^\infty k\: dk \:
n^A(k)
\eeq
 is the longitudinal momentum distribution, and the second
one
\beq \label{binding}
B^A(q,Y)=~2\pi \: \int_{E_{min}}^\infty dE \:
\int_{|Y|}^{k_{min}(q,Y,E)} k\:dk\: P^A_1(k,E)
\eeq
 is the so called "binding correction".
\begin{figure}[!h]
\centerline{\centerline{\epsfig{file=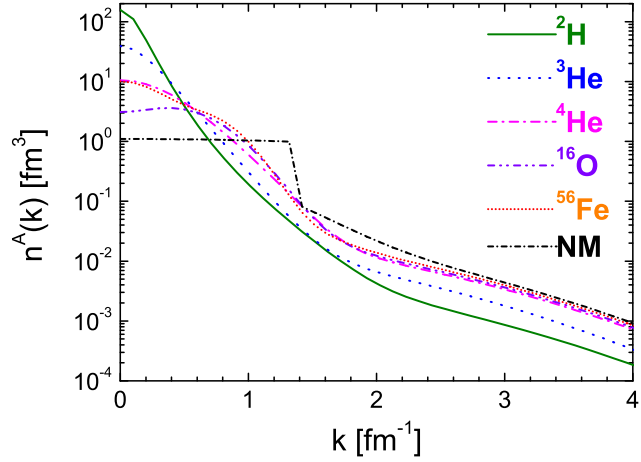,width=10.cm,height=7cm}}}
\vskip -0.3cm \caption{The nucleon momentum distributions $n^A(k)$ for
nuclei ranging from $^2H$ to $NM$.
It can be seen that, at high values of the momentum $k$, $n^A(k)$ can be considered as a
rescaled version of the momentum distributions of $^2H$. After Ref. \cite{ciosim}.}\label{Fig2}
\end{figure}
The longitudinal momentum distribution depends only upon the
nucleon momentum distributions $n^A(k)=\int P^A(k,E)\:dE$, which, as is well known
\cite{urbana} and  illustrated in
Fig. \ref{Fig2}, at high values of the momentum $k$ scale with A
according to $n^A(k) \simeq C^A\:n^D(k)$; the binding correction
$B^A(q,Y)$, on the contrary, depends upon the correlated part of
the spectral function $P^A_1(k,E)$
(as is well known, $P^A(k,E)=P_0^A(k,E)+P_1^A(k,E)$,
where $P_0^A(k,E)$ is the
  (trivial) shell-model part
and $P_1^A(k,E)$ is the (interesting) part generated by NN correlations \cite{simo}). In the Deuteron case, one has
$E=E_{min}=2.22\: MeV$, $k_{min}(q,Y,E_{min})=|Y|$, $B^D(q,Y)=0$
and $F^D(q,Y)=f^D(Y)$, from which the nucleon momentum
distributions can be obtained by the relation $n^A(k)=-[df^A(Y)/dY]/[2\pi Y]$;
in general, however, $B^A(q,Y) \neq 0$ and $F^A(q,Y) \neq f^A(Y)$
and the momentum distributions cannot be obtained. The central idea
of our approach \cite{ciocbm}, is that the contribution arising
from the binding correction could be minimized by a proper choice
of the scaling variable $Y$, such that  $k_{min}(q,Y,E)\simeq
|Y|$, with the resulting  cross section (\ref{X-section}) depending
only upon the nucleon momentum distributions, obtaining, by this
way, a direct access  to high momentum components generated by SRC. It is clear
that the outlined picture can in principle be modified by the effects of the final state
interactions (FSI); this important point will be discussed later on.
%
%
\subsection{Traditional approach to Y-scaling: the mean field scaling variable}
The traditional scaling variable, usually denoted by small $y$,
 $Y
\equiv y$, is obtained by placing $k=|y|$,
$\cos\alpha=(\textbf{k}\cdot \textbf{q}/kq)= 1$ and $E_{A-1}^*=0$
in the energy conservation law given by \beq \label{energy}
\nu+M_A=\sqrt{(M_{A-1}+E_{A-1}^*)^2+\textbf{k}^2}+\sqrt{m_N^2+(\textbf{k}+\textbf{q})^2}
\eeq where $E_{A-1}^*$ is the intrinsic excitation energy of the
$(A-1)$-nucleon system and the other notations are self explained.  In
such an approach, $y$ represents the minimum longitudinal momentum
of a nucleon having the minimum value of the removal energy
$E=E_{min}+E_{A-1}^*=E_{min}=m_N+M_{A-1}-M_A$.
\begin{figure}[!h]
\centerline{\centerline{\epsfig{file=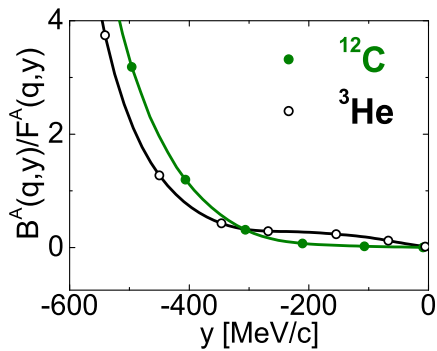,width=7.0cm,height=5cm}}}
\vskip -0.3cm \caption{The ratio of the binding correction $B^A(q,y)$ (Eq. (\ref{binding})) to the scaling function $F^A(q,y)$ (Eq. (\ref{Funzscala})) for $^3He$ (open dots) and $^{12}C$ (full dots), calculated using the scaling variable $y$. After Ref. \cite{ciocbm}.}\label{Fig3}
\end{figure}
\begin{figure}[!h]
\centerline{\centerline{\epsfig{file=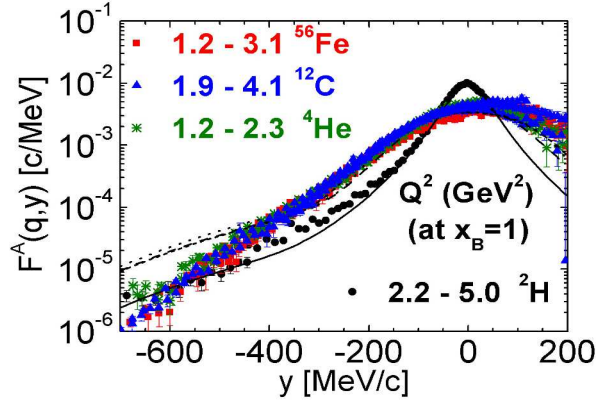,width=8.0cm,height=5.5cm}}}
\vskip -0.3cm \caption{The experimental scaling function $F^A_{exp}(q,y)$
of $^4He$,
  $^{12}C$, and $^{56}Fe$ obtained from the experimental data of
  Refs. \cite{deuteron,arrington}. The longitudinal momentum
  distributions (Eq. (\ref{longitudinal})) of $^{2}H$ (full line),
  $^4He$ (long-dashed),  $^{12}C$ (dashed) and  $^{56}Fe$ (dotted) are also shown. After Ref. \cite{ciocbm}.}\label{Fig4}
\end{figure}
In the asymptotic limit, $(q \rightarrow \infty)$, one has
 $k_{min}^\infty(q,y)=|y-(E-E_{min})|$ \cite{ciofi},
so that, when
$E=E_{min}$, $k_{min}^\infty(q,y)=|y|$  and $B^A(q,y)=0$; this occurs only when $A=2$, whereas in the general case,
$A>2$,  the excitation energy  $E_{A-1}^*$ of the
residual system is different from zero, leading to $B^A(q,y)>0$.
The binding correction plays indeed a relevant role in the traditional approach to $Y$-scaling. To illustrate this,
the ratio $B^A(q,y)/F^A(q,y)$ is shown in Fig. \ref{Fig3}; it can be
seen that at high (negative) values of $y$, the effects from binding are very
large. Moreover, the experimental scaling function
$F_{exp}^A(q,y)=\sigma_{exp}/[K(q,y)\,\:(Z\sigma_{ep}+N\sigma_{en})]$
plotted versus the scaling variable $y$
confirms, as shown in Fig. \ref{Fig4}, that the scaling function strongly differs from the
longitudinal momentum distribution, and therefore does not exhibits any
proportionality to the Deuteron scaling function $f^D(y)$.
%
%
\subsection{A novel approach to $Y$-scaling: the scaling variable embedding two-nucleon correlations}
\begin{figure}[!h]
\centerline{\centerline{\epsfig{file=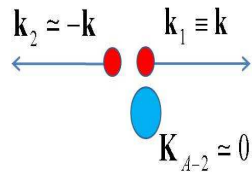,width=3.4cm,height=2.5cm}}}
\vskip -0.3cm \caption{2NC correlations in a nucleus A: the high
momentum $\textbf{k}_1 \equiv \textbf{k}$ of  nucleon "1" is almost completely balanced
by the momentum $\textbf{k}_2\simeq -\textbf{k}$ of the partner nucleon "2",
 whereas the residual system moves with low momentum $\textbf{K}_{A-2}$. Momentum conservation is $\sum_1^A\:\textbf{k}_i=\textbf{k}_1+\textbf{k}_2+\textbf{K}_{A-2}=0$.}\label{Fig2NC}
\end{figure}

\begin{figure}[!h]
\centerline{\centerline{\epsfig{file=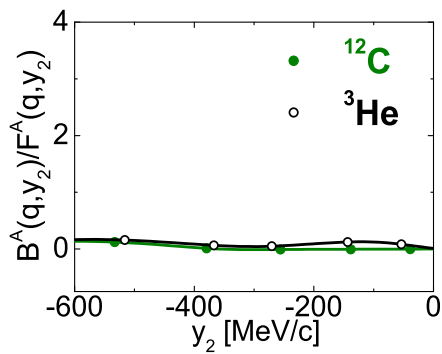,width=7.0cm,height=5cm}}}
\vskip -0.4cm \caption{The same as in Fig. \ref{Fig3}, obtained using
 in Eqs. (\ref{Funzscala}) and (\ref{binding}) the scaling variable $y_{2}\equiv y_{CW}$. After Ref. \cite{ciocbm}.}
\vskip 0.5cm
\label{Fig5}
\end{figure}

2NC are defined as those nucleon configurations shown
in Fig. \ref{Fig2NC} \cite{ratioAD}: momentum conservation in the ground state of the target nucleus $(\sum_{1}^A \textbf{k}_i=0)$  is almost entirely exhausted by two correlated nucleons with high momenta,
 the $(A-2)$-nucleon system acting mainly as a spectator, moving with very low momentum.
The intrinsic excitation energy of the $(A-1)$-nucleon system is in this case
\beq
E_{A-1}^*=\frac{(A-2)}{(A-1)}\:\frac{(\textbf{k}_2-\textbf{K}_{A-2})^2}{2m_N}
\eeq
which becomes
\beq
E_{A-1}^*=\frac{(A-2)}{(A-1)}\:\frac{k^2}{2m_N}
\eeq
in the naive 2NC model, i.e. the model based upon the assumption $\textbf{K}_{A-2}=0$.
Since high excitation states of the final
$(A-1)$-nucleon system are generated by SRC in the ground state of the target nucleus, the traditional (mean field) scaling variable
$y$ does not incorporate, by definition, SRC effects, for it is
obtained by placing $E_{A-1}^*=0$ in the energy conservation law
(\ref{energy}). Motivated by this observation, in Ref. \cite{CW},
a new scaling variable $Y\equiv y_{CW} \equiv y_2$ has been
introduced by setting in Eq.
(\ref{energy}) $k=|y_2|$, $\cos\alpha=(\textbf{k}\cdot
\textbf{q}/kq)= 1$ and $E_{A-1}^*=< E_{A-1}^* (k)>_{2NC}$.
 By this way,  $y_2$ properly includes the momentum
dependence of the average excitation energy of the $(A-1)$-nucleon
system generated by SRC. The approach of Ref. \cite{CW} has been
further improved in Ref. \cite{ciocbm}, obtaining a
scaling variable $y_2$ which, through the $k$
dependence of $<E_{A-1}^*(k)>_{2NC}$,  interpolates between the correlations and
the mean field regions of the q.e. cross section.
The relevant feature of $y_2$ is that it leads to
$k_{min}(q,y_2,E)\simeq~|y_2|$ and therefore to a minor role of the
binding correction; this is indeed demonstrated in Fig. \ref{Fig5}, which clearly shows that
$B^A(q,y_2)$ vanishes in the whole region of $y_2$ considered. One
can therefore conclude that, using the new scaling variable, one obtains $F^A(q,y_2) \sim f^A(y_2)\sim C^A \:
f^D(y_2)$.
\begin{figure}[!h]
\centerline{\centerline{\epsfig{file=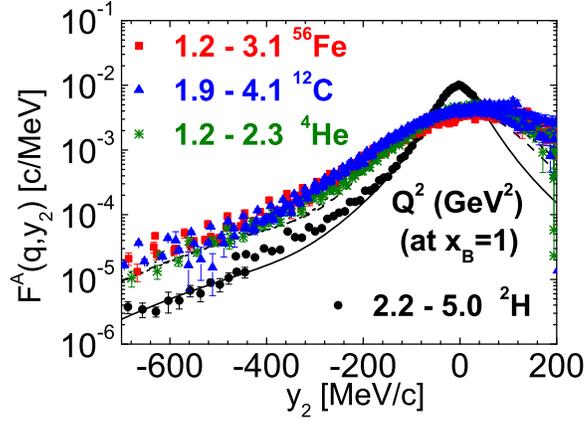,width=8.5cm,height=6cm}}}
\vskip -0.4cm \caption{The same as in Fig. \ref{Fig4} \emph{vs.} the scaling variable
$y_{2}\equiv y_{CW}$. After Ref.
\cite{ciocbm}.}
\vskip 0.5cm
\label{Fig6}
\end{figure}
%
%
\\The scaling function $F^A(q,y_2)$ obtained from available
 experimental data  on $^4He$, $^{12}C$ and $^{56}Fe$ is
plotted in Fig. \ref{Fig6} versus the scaling variable $y_2$;
it can be seen that at high values of $|y_2|$, the relation $F^A(q,y_2) \sim
f^A(y_2)\sim C^A \: f^D(y_2)$ is indeed experimentally confirmed. In
order to analyze more quantitatively the scaling behavior of
$F^A(q,y_2)$, the latter has been plotted versus $Q^2$, at fixed values of $y_2$.
\begin{figure}[!h]
\centerline{\centerline{\epsfig{file=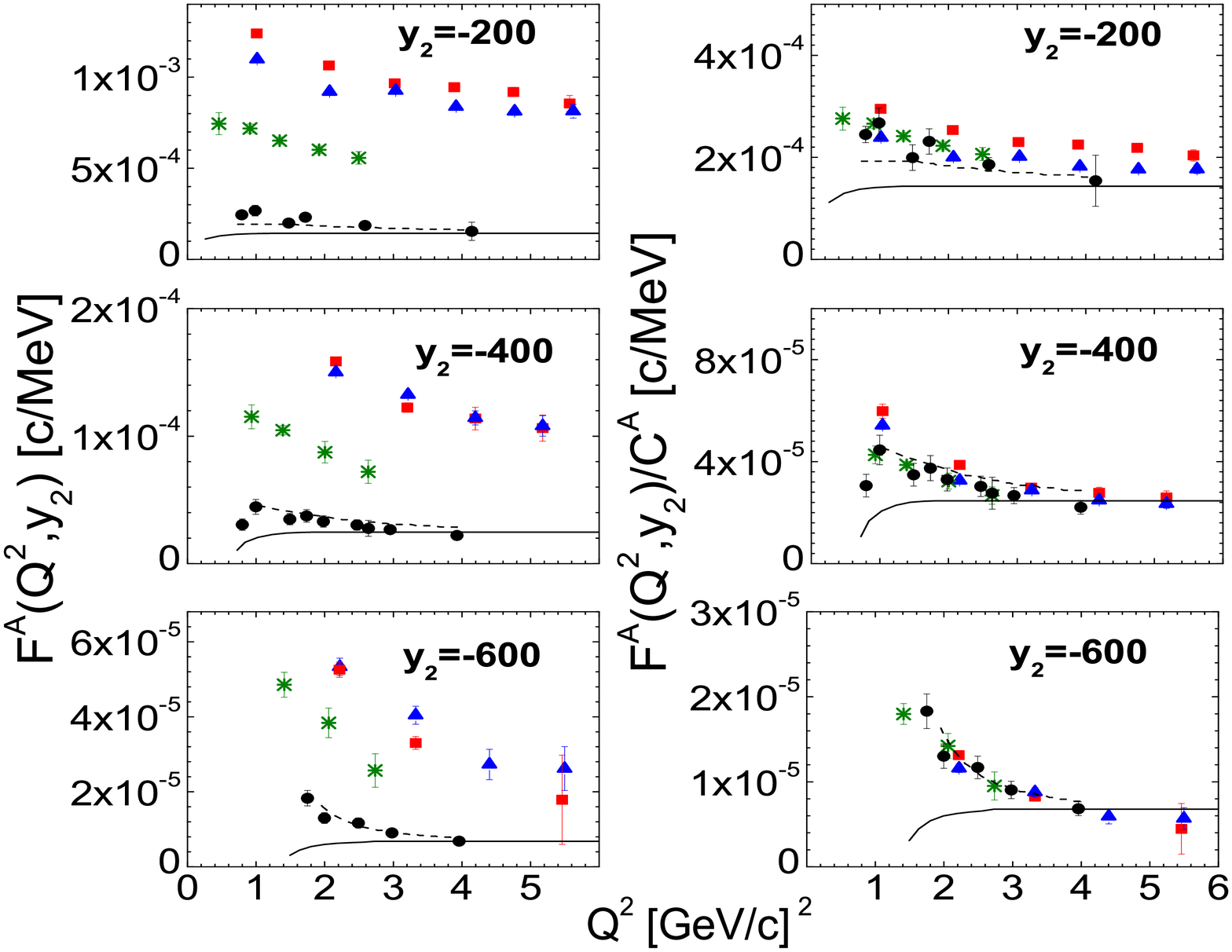,width=10.0cm,height=9.0cm}}}
\vskip -0.4cm \caption{{\it Left panel}: the scaling
function $F^A(Q^2,y_{2})$ \emph{vs.} $Q^2$, at fixed values of $y_2\equiv y_{CW}$.
{\it Right panel}: the same data divided by the constants $C^4=2.7$, $C^{12}=4.0$
and $C^{56}=4.6$, respectively for $^4He$, $^{12}C$ and $^{56}Fe$. The theoretical curves
 represent the longitudinal momentum of the Deuteron,  calculated (AV18 interaction)  in PWIA (full line) and including FSI (dashed line) effects. After Ref. \cite{ciocbm}.}\label{Fig7}
\vskip 0.5cm
\end{figure}
The result is shown in Fig. \ref{Fig7}, together with the theoretical scaling function for $A=2$, calculated in PWIA (solid line), and taking FSI into account (dashed line).
It can be seen (left panel) that, due to FSI effects, scaling is violated and approached  from the top,
and not from the bottom, as predicted by the PWIA. However, the
 violation of scaling seems to exhibit a $Q^2$ dependence  which is very
similar in Deuteron and in complex nuclei. This is illustrated in
more details in the right panel of the figure, which shows
$F^A(Q^2,y_2)$  divided by a constant $C^A$, chosen so as to
obtain the Deuteron scaling function $F^D(Q^2,y_2)$. It clearly appears that the scaling function of heavy and light nuclei
 scales to the Deuteron scaling function; moreover the values obtained for $C^A$ turn out to
be in agreement with the ones predicted in Ref. \cite{ratioAD}; it is also important to stress that,
although FSI are very relevant, they appear to be similar in Deuteron and in
a nucleus A, which is evidence that, in the SRC region, FSI are
mainly restricted to the correlated pair.
%
%
\subsection{A novel approach to $Y$-scaling: the scaling variable embedding three-nucleon correlations}
\begin{figure}[!h]
\centerline{
    \epsfig{file=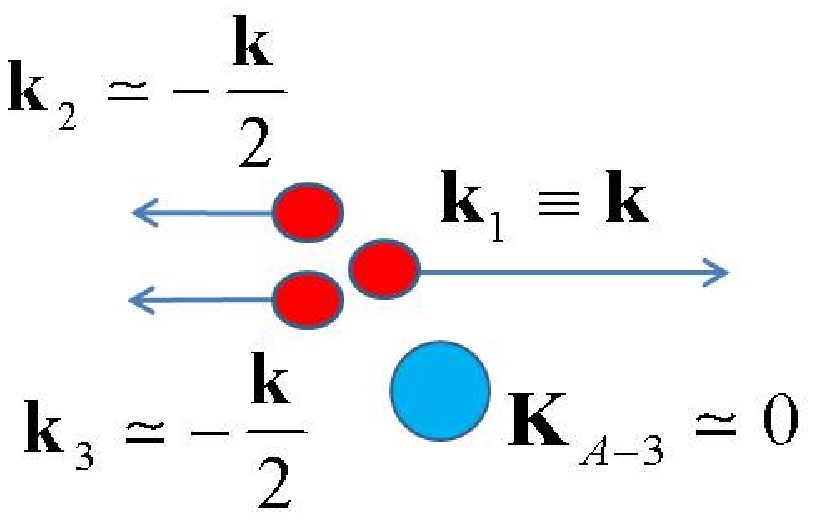,width=3.cm,height=3.cm}
    \hspace{0.8cm}
    \epsfig{file=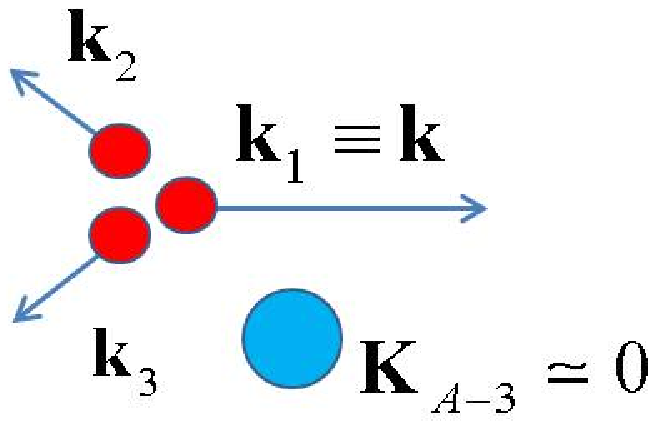,width=3.cm,height=3cm}}
    \vskip -0.5cm
\centerline{$(a)$\hspace{3.3cm}$(b)$}
\vskip -0.4cm \caption{Two types of  3NC  configurations which are present in the
spectral function
of a nucleus A; they  correspond to: (a)
high momentum $k$ and low removal energy $E$, and (b) to high momentum
$k$ and high removal energy $E$.}\label{3NC}
\vskip 0.5cm
\end{figure}

\begin{figure}[!h]
\centerline{\centerline{\epsfig{file=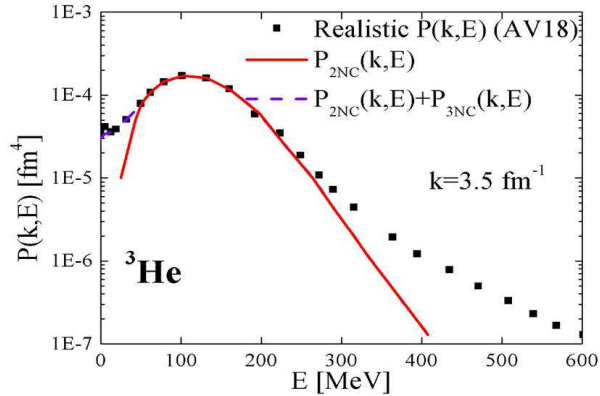,width=8.0cm,height=5.5cm}}}
\vskip -0.5cm \caption{The  spectral function of $^3He$ \emph{vs.} the
removal energy $E$, at $k=3.5\: fm^{-1}$ \cite{CK}.
The realistic spectral function corresponding to the Pisa wave
function (squares) is compared with the  2NC model of
Ref. \cite{ciosim} (full line) and with a model which includes also
3NC (dashed line) of the type depicted in Fig. \ref{3NC}(a) \cite{ciocbm2}.}\label{Fig8}
\vskip 0.5cm
\end{figure}

\begin{figure}[!h]
\centerline{\centerline{\epsfig{file=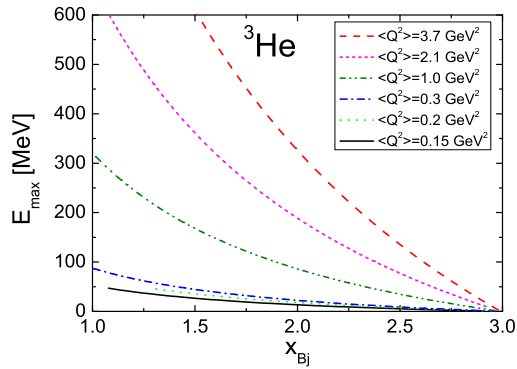,width=8.0cm,height=5.5cm}}}
\vskip -0.4cm \caption{The maximum value of the removal energy $E_{max}$ (Eq. (\ref{emax}))
available in inclusive q.e. scattering off $^3He$ plotted\emph{ vs.} $x_{Bj}$ at increasing
values of $Q^2$, shown in the inset.}\label{Fig9}
\vskip 0.5cm
\end{figure}

Let us now consider three-nucleon correlations. These correspond to those
three-nucleon configurations when the high momentum $\textbf{k}_1 \equiv \textbf{k}$ of nucleon "1" is almost entirely balanced by the momenta $\textbf{k}_2$ and $\textbf{k}_3$ of nucleons "2" and "3". The excitation energy of the $(A-1)$-nucleon system is given in this case by
\beq \label{Energy3NC}
E_{A-1}^*=\frac{\left(\textbf{k}_2-\textbf{k}_3\right)^2}{m_N} + \frac{A-3}{A-1}\:\frac{\left[\left(\textbf{k}_2+\textbf{k}_3\right)-2\textbf{K}_{A-3}\right]^2}{4m_N}
\eeq
Eq. (\ref{Energy3NC}) shows that, whereas 2NC are
directly linked to high values of excitation energies $E_{A-1}^*
\simeq (A-2)\: k^2/2m_N(A-1)$, high momentum components due to
3NC may lead both to low and to high values of $E_{A-1}^*$, as shown in the examples of Figs. \ref{3NC}(a)
 and \ref{3NC}(b), respectively.
In the configuration of Fig. \ref{3NC}(a),
the momentum $\textbf{k}_1 \equiv \textbf{k}$ of nucleon $"1"$
is almost entirely balanced by nucleons $"2"$ and $"3"$, with momenta
$\textbf{k}_2\simeq\textbf{k}_3\simeq-\textbf{k}/2$, and one has
\beq
E_{A-1}^*=\frac{A-3}{A-1}\:\frac{k^2}{4m_N}
\eeq
In the configuration of Fig. \ref{3NC}(b), $k_2=k_3=-|\textbf{k}|\cos(\theta/2)/2$,  with
$\cos\theta=-(\textbf{k}_2\cdot \textbf{k}_3)/(k_2 k_3)$, and
$E_{A-1}^*$ could be very large.
Let us investigate the presence and relevance of 3NC configurations in the spectral function of
 the 3-nucleon system for which the Schroedinger equation has been
 solved exactly. When $A=3$, 3NC of the type shown in Fig.
 \ref{3NC}(a) lead to  $E_2^*=0$. In Fig. \ref{Fig8}, the realistic spectral function of
$^3He$ obtained \cite{CK} using the Pisa wave function \cite{pisa}
corresponding to the AV18 interaction \cite{av18} (full squares),
is compared with the predictions of the 2NC model (solid line)
\cite{ciosim} and with a model which includes also 3NC of the type
depicted in Fig. \ref{3NC}(a) (dashed line) \cite{ciocbm2}. It can be
observed that 2NC reproduce the exact spectral function in a wide
range of removal energies ($50 \lesssim E \lesssim 200\: MeV$),
but fail at very low and very high values of E, where the effects
from 3NC are expected to provide an appreciable contribution.
It is clear  from Fig. \ref{Fig8} that 3NC of the type shown in Fig. \ref{3NC}(b) can
 hardly be present in the spectral function at $k<3.5 fm^{-1}$ and  $E \leq 300\: MeV$,
 so that it is legitimate  to ask ourselves whether these 3NC can show up in available
 experimental data. To answer this question, let us now consider the maximum value of the
 removal energy achieved in the experiments, i.e.  the upper limit of integration
  in Eq. (\ref{Funzscala}),
\beq \label{emax}
E_{max}(q,\nu)=\sqrt{(\nu+M_A)^2-q^2}
 \eeq
In Fig. \ref{Fig9},  we show  $E_{max}(q,\nu)$ plotted versus the Bjorken scaling variable
in the region $1 \leq x_{Bj} \leq 3$ in correspondence of a set of values
of $\nu$ and $q$ typical of available experimental data on $^3He$.
It can be seen from Figs. \ref{Fig8} and \ref{Fig9}
that in the region $ 2\leq x_{Bj} \leq 3$ only  3NC configurations of the type shown in
 Fig. \ref{3NC}(a) can contribute
to present $A(e,e')X$
 kinematics;  for this reason we will consider, for
 the time being, only this type of 3NC .

%
%
The scaling variables $y$ and $y_2$ have been obtained by placing
 different values of $E_{A-1}^*$ in Eq. (\ref{energy}), namely
 $E_{A-1}^*=0$ and $E_{A-1}^*=<E_{A-1}^*(k)>_{2NC}$, respectively.
 Following the same procedure,  we have derived the scaling variable embedding
\begin{figure}[!h]
\centerline{\centerline{\epsfig{file=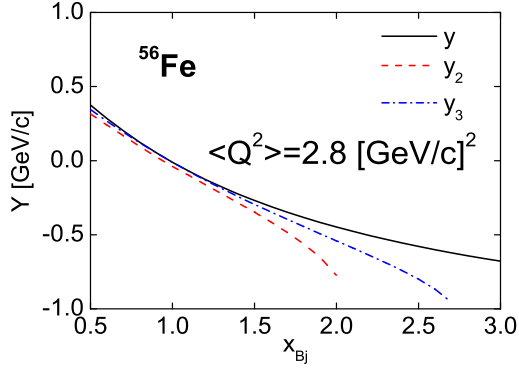,width=8.0cm,height=5.5cm}}}
\vskip -0.4cm
\caption{The scaling variables $y$, $y_2$ and $y_3$ {\it vs.} $x_{Bj}$ for  $A=56$.}\label{Fig10}
\vskip 0.5cm
\end{figure}
 3NC, $Y \equiv y_3$, by placing in Eq. (\ref{energy})
  $E_{A-1}^*=<E_{A-1}^*(k)>_{3NC}$,
 with $<E_{A-1}^*(k)>_{3NC}$ calculated within the 3NC configuration \ref{3NC}(a),
  which corresponds to high values of $k$ and small values of $E$.
  The explicit expression for $y_3$ will be given elsewhere \cite{ciocbm2}.
  Here we show in Fig. \ref{Fig10}, in the case of $^{56}Fe$, the values of
  $y$, $y_2$ and $y_3$ plotted versus $x_{Bj}$. It can be seen that, because of the different
  values of $E_{A-1}^*$ used in Eq. (\ref{energy}), different limits of existence of
  the three scaling variables are obtained: $y$ describes
the mean field configuration and is defined in the whole range of $x_{Bj}\leq A$; $y_2$
represents 2NC in heavy nuclei resembling the ones acting in Deuteron and is defined
only for $x_{Bj}\leq 2$; $y_3$, eventually, describes 3NC as in $^3He$, and is defined only
for values of $x_{Bj}$ up to $3$.
%
%
\section{Cross section ratio: Preliminary results}
As mentioned in previous sections, our novel approach to inclusive
 lepton scattering off nuclei is based upon the introduction of proper scaling variables
  that effectively include the energy $E_{A-1}^*$ of the residual system and allow one to
  describe the $A(e,e')X$ cross section only in terms of nucleon momentum distributions
  generated by 2N and 3N SRC, i.e.
\bey
  &&\frac{d^2\sigma}{d\Omega_2\:d\nu}\propto \int_{E_{min}}^{E_{max}(q,\nu,E)}
  dE \int_{k_{min}(q,\nu,E)}^{k_{max}(q,\nu,E)} kdk \: P^A(k,E) \nonumber \\
  &\simeq&  \int_{|y|}^\infty n_0^A(k)\: k dk +\int_{|y_2|}^\infty n_2^A(k)\: k dk +\int_{|y_3|}^\infty n_3^A(k)\: k dk \nonumber \\
\eey
where $n_0^A(k)$ is the component of the nucleon momentum distribution generated by the mean field,
\beq \label{soft}
    n_2^A(k)=\int dk_{CM}\: n_{rel}(\textbf{k}+\textbf{k}_{CM})\:n_{CM}^{soft}(\textbf{k}_{CM})
\eeq
is the one due to 2NC and, eventually,
\beq \label{hard}
n_3^A(k)=\int dk_{CM}\: n_{rel}(\textbf{k}+\textbf{k}_{CM})\:n_{CM}^{hard}(\textbf{k}_{CM})
\eeq
is the one due to  3NC; here, $n_{CM}^{soft}$ and $n_{CM}^{hard}$ include only
"soft" and "hard" momentum components, respectively.
Within such an approach, the cross section ratio $r(A/A')$
reduces to the scaling function ratio of nuclei $A$ and $A'$.
\begin{figure}[!h]
\centerline{\centerline{\epsfig{file=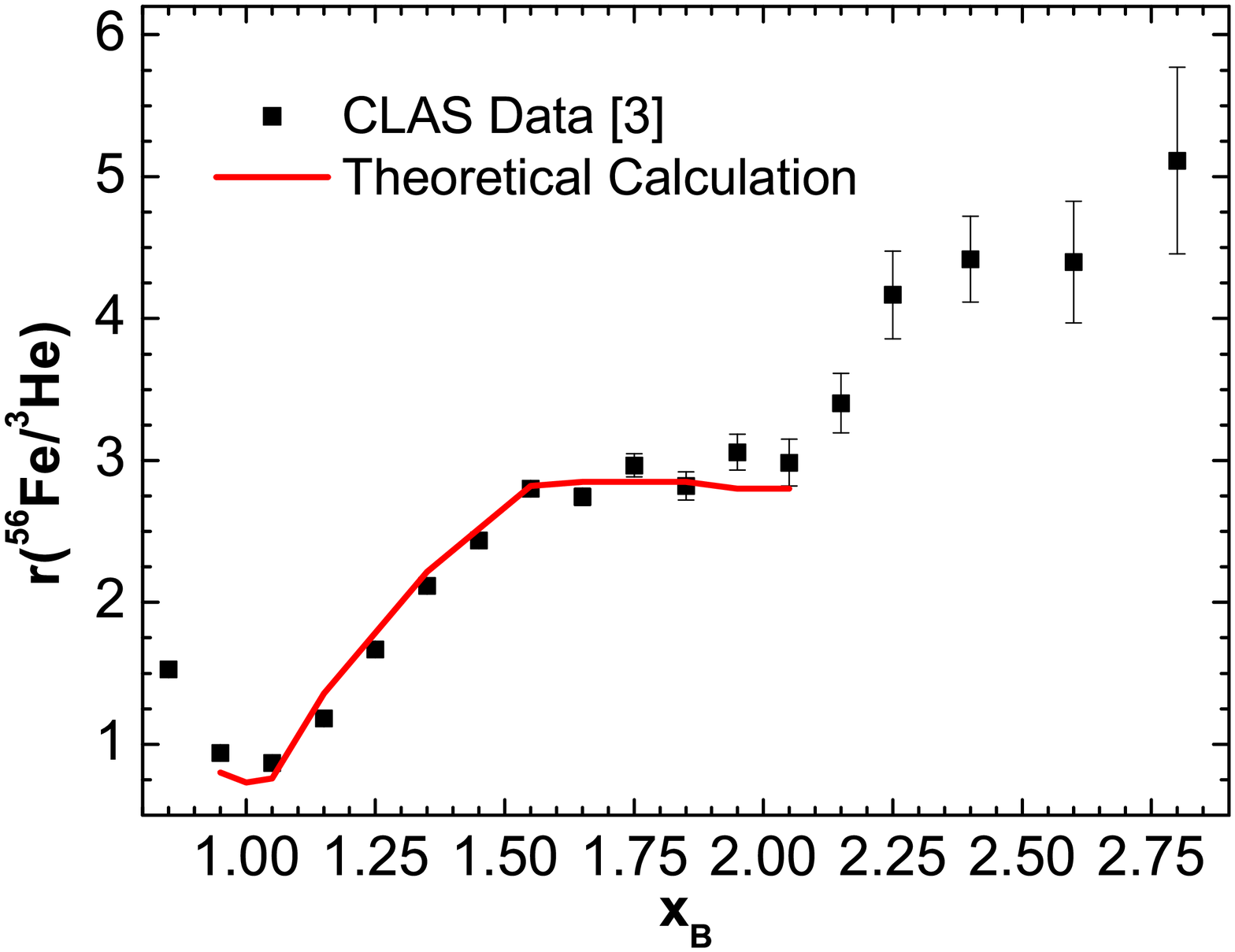,width=8.5cm,height=7.0cm}}}
\vskip -0.4cm \caption{The experimental cross section ratio shown in Fig. \ref{Fig1} compared with our preliminary theoretical results.}\vskip 0.5cm
\label{Fig11}
\end{figure}
Our preliminary calculations of the scaling function ratio, performed for $A=56$, show in PWIA a good agreement with CLAS data only for $1.5 \lesssim x_{Bj}\lesssim 2$, i.e. in the region of 2NC; on the contrary, at $x_{Bj}\lesssim 1.5$, the PWIA does not lead to satisfactory results. This fact agrees with the results already shown in Fig. \ref{Fig7}: in the region of 2NC the data of heavy nuclei scale to the Deuteron ones, and thus FSI effects vanish in the ratio $r(A/A')$, leading to the first plateaux; in the kinematical region at $x_{Bj}\lesssim 1.5$, on the opposite, the ratio exhibits a strong sensitivity upon the A-dependent FSI of the knocked nucleon with the residual system. Including explicitly these FSI effects in the mean field contribution, we obtained the preliminary results shown in Fig. \ref{Fig11}. Calculations of 3NC effects in the region $2 \lesssim x_{Bj} \lesssim 3$ are in progress and will reported elsewhere \cite{ciocbm2}.
%
%
\section{Conclusions}
To sum up, the following remarks are in order:
\begin{itemize}
\item the experimental scaling function in the 2NC region scales to the Deuteron scaling
function and exhibits A-independent FSI effects, mostly due to the FSI in the correlated pair;
\item proper scaling variables have been introduced which effectively
include the excitation energy $E_{A-1}^*$ of the residual system in different ways and allow one
to describe the $A(e,e')X$ cross section in terms of
the corresponding momentum distributions generated by  2NC and 3NC;
\item  the experimental ratio $r(^{56}Fe/^3He)$ in the 2NC region has been successfully reproduced.
\end{itemize}

 Calculations including the 3NC configurations of
the type shown in Fig. \ref{3NC}(b), which are necessary
in order to extend our comparison
with the CLAS experimental data to the region $2 \lesssim x_{Bj} \lesssim 3$,  are in progress and the explicit
 introduction of FSI within the correlated pair is being considered \cite{ciocbm2}.

\end{document}